\documentclass[reprint,
 amsmath,amssymb,
 aps,
 prc
]{revtex4-2}

\usepackage{graphicx}
\usepackage{dcolumn}
\usepackage{bm}
\usepackage{epstopdf}
\usepackage[dvipsnames]{xcolor}

\begin{document}

\preprint{APS/123-QED}

\title{Analog-antianalog isospin mixing in $^{47}$K $\beta^-$ decay}

\author{Brian Kootte$^1$}
\author{H. Gallop$^{1,3}$}%
\author{C. Luktuke$^{1,3}$}
\author{J.C. McNeil$^{2,1}$}
\author{A. Gorelov$^1$}
\author{D.G. Melconian$^{4,5}$}
\author{J. Klimo$^{4,5}$}
\author{B.M. Vargas-Calderon$^{4,5}$}
\author{J.A. Behr$^{1,2}$}
\email{behr@triumf.ca}
\affiliation{$^1$TRIUMF, 4004 Wesbrook Mall, Vancouver, BC V6T 2A3 Canada\\
$^2$ University of British Columbia, Department of Physics and Astronomy, 6224 Agricultural Road, Vancouver, B.C. V6T 1Z1 Canada\\
$^3$University of Waterloo, Department of Physics and Astronomy, 200 University Ave W, Waterloo, ON N2L 3G1\\
$^4$ Cyclotron Institute, Texas A\&M University, 3366 TAMU, College Station, Texas 77843-3366, USA\\
$^5$ Department of Physics and Astronomy, Texas A\&M University,
4242 TAMU, College Station, Texas 77843-4242, USA
}




\date{\today}

\begin{abstract}
We have measured the isospin mixing of the I$^\pi$= 1/2$^+$ E$_x$=2.599 MeV state in nearly-doubly-magic $^{47}$Ca with the isobaric analog 1/2$^+$ state of $^{47}$K.
Using the TRIUMF atom trap for $\beta$ decay, we have measured a nonzero asymmetry of the progeny $^{47}$Ca with respect to the initial $^{47}$K spin polarization, which together with the $\beta$ asymmetry implies a nonzero ratio of Fermi to Gamow-Teller matrix elements $y$= 0.098$\pm$0.037 for the $1/2^+ \rightarrow 1/2^+$ transition. 
Interpreting $y$ as mixing between this state and the isobaric analog state implies a Coulomb matrix element magnitude 101 $\pm$ 37 keV.
This relatively large matrix element supports a model from the literature of analog-antianalog isospin mixing, which predicts large matrix elements in cases involving excess neutrons over protons occupying more than one major shell.
%
The result supports pursuing a search for
time-reversal odd, parity-even, isovector interactions using a correlation in $^{47}$K $\beta$ decay.
\end{abstract}

\maketitle


\section{\label{sec:level1}Introduction} 

The neutron beta decays to its isobaric analog state, the proton, as does tritium.
Many other isotopes undergo beta minus decay to states of same spin $I$ and parity $\pi$,
but because of the extra Coulomb energy at higher Z, decay to the isobaric analog state is
energetically forbidden. So the Gamow-Teller operator dominates, while the Fermi operator
linking isobaric analog states is only allowed if some low-lying final state of same $I^\pi$
is mixed by an isospin-breaking interaction with the excited isobaric analog.
We see such isospin breaking in an $I^\pi$=1/2$^+$ state in the $^{47}$Ca nucleus 80\% fed by the beta decay of $^{47}$K. Interference between Gamow-Teller and isospin-suppressed Fermi amplitudes
produces an asymmetry of the progeny recoil direction with respect to the initial nuclear spin, which we measure with TRIUMF's Neutral Atom Trap for $\beta$ decay (TRINAT).

Since $^{47}$Ca and $^{47}$K are near closed shells, the single known $^{47}$Ca 1/2$^{+}$ state may contain much of the antianalog configuration, 
and is predicted to have a relatively large Coulomb mixing matrix element with the analog~\cite{Auerbach2022}.
Sensitivity to time reversal-odd parity-even (TOPE) inherently isovector~\cite{Simonius1997}
N-N interactions through a $\beta$-$\nu$-spin correlation is thought to be enhanced in these systems, as the small amount of time reversal is referenced to Coulomb rather than strong interactions~\cite{Barroso1973},
which motivates our measurement of isospin breaking in $^{47}$K decay.

\section{Theory and Methods}

\subsection{Analog $\mathcal{A}$ - antianalog $\bar{\mathcal{A}}$ mixing}




The antianalog configuration has same spin and occupancy of spatial orbitals as the isobaric analog, but has total isospin reduced to $T=T_z$, with the antisymmetry of its wavefunction encoded differently between spin and isospin parts so that it
is orthogonal to the analog state.
Auerbach and Loc~\cite{Auerbach2022}, using schematic wavefunctions, write
for analog-antianalog Coulomb mixing for $n_1$ and $n_2$ excess neutrons over protons occupying orbitals $j_1$ and $j_2$:

\begin{eqnarray}
H_C =    \langle \bar{\mathcal{A}} | V_{\rm C} | \mathcal{A} \rangle & = \frac{\sqrt{n_1 n_2}}{2T} (\langle j_1 | V_C |j_1 \rangle - \langle j_2 | V_C |  j_2\rangle) \\
%
 & \rightarrow 0.35 \frac{\sqrt{n_1 n_2}}{2T} \frac{Z}{A^{2/3}} {\rm MeV},
\end{eqnarray}
for harmonic oscillator wavefunctions, a uniform charge distribution, and cases where
%
neutrons occupy two major shells differing by one $\hbar \omega$.
Ref.~\cite{Auerbach2022}, using experimental Coulomb energies from mirror nuclei, note that Eq.~1 produces for occupancy of both the $p_{3/2}$ or $p_{1/2}$ and $f_{5/2}$ or $f_{7/2}$ subshells $H_C$ about a factor of two smaller than the major shell-occupancy Eq.~2, thus $f_{7/2}$ is not a major shell when considering Coulomb energies. 
Ref~\cite{Auerbach2022}  then benchmark these simple expressions with RPA calculations to an accuracy of $\sim$20\%.  
Eq.~2 predicts $H_C$ = 160 keV for our case   $^{47}_{19}$K$^{28}$ (with $n_1$=8 f$_{7/2}$ neutrons and $n_2$=1 2s$_{1/2}$ neutron excess over protons).

\subsection{Isospin-suppressed $\beta$ decay}

In the angular distribution for allowed I=1/2 $\beta$ decay in terms of lepton momentum $p$ and energy $E$ ~\cite{Jackson1957,Holstein1974}:

\begin{equation}
dW = F(E,Z) p E p_\nu E_\nu (1 + a \frac{\vec{p_\beta}}{E_\beta} \cdot \frac{\vec{p_\nu}}{E_\nu}+\hat{I} \cdot (A_\beta \frac{\vec{p_\beta}}{E_\beta} + B_\nu \frac{\vec{p_\nu}}{E_\nu})),
\end{equation}

\noindent isospin-suppressed Fermi decay alters the correlation coefficients from their Gamow-Teller values:

\begin{equation}
a= \frac{y^2-1/3}{y^2+1}; A_\beta =A_{\beta {GT}} + f(M_{F}) ; B_\nu = -A_{\beta {GT}} +f(M_F)
\end{equation}
with  $y=g_V M_F/g_A M_{GT}$ and 
   $f(M_F)=2 \sqrt{\frac{I}{I+1}}\frac{y}{1+y^2}$.
The recoil asymmetry $A_{\rm recoil}$ is then proportional to $A_{\beta}$+$B_{\nu}$, 
which vanishes when $M_F$=0. (Analytic expressions for the proportion, which are possible if the Fermi function is set to unity, are given in Refs.~\cite{Pitcairn2009,Treiman1958}:  we compare here entirely to numerical simulations.)

\subsection{$^{47}$K and its $\beta^-$ decay to $^{47}$Ca }
\begin{figure}[htb]
    \centering
  \includegraphics[width=0.7\linewidth]{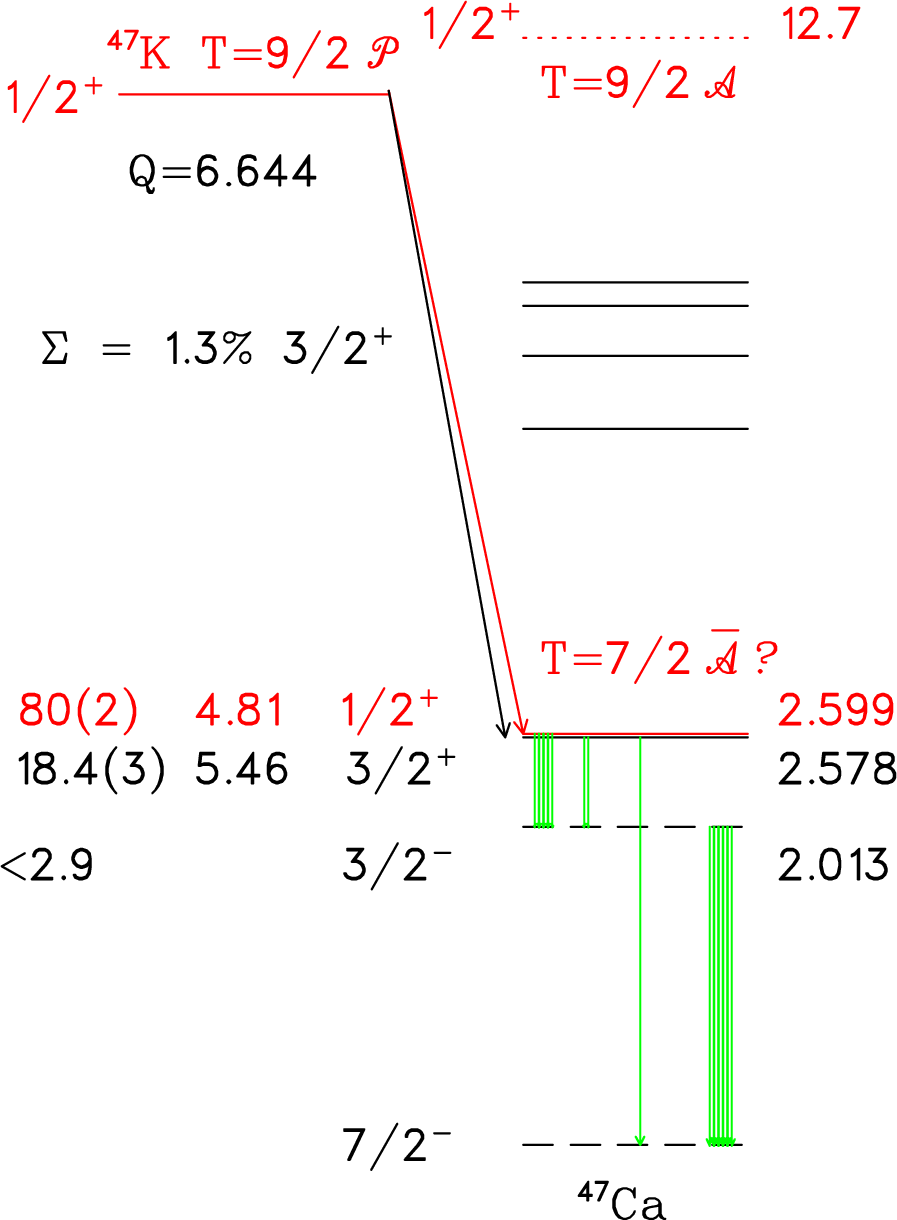}
    \caption{
    Relevant allowed decay of $^{47}$K, showing $\beta^-$ branches $>$0.04\%, log(ft),  $I^\pi$, energy [MeV], and isospin $T$ of the isobaric parent $\mathcal{P}$, analog $\mathcal{A}$, and possible antianalog $\bar{\mathcal{A}}$. Thickness of
    $\gamma$ transitions $>$5\% indicate intensity.
    }
    \label{fig:47Kdecay}
\end{figure}

The $^{47}$K I$^\pi$=1/2$^+$ ground-state has magnetic moment  1.933(9) $\mu_N$ = 0.69 $\mu_{\rm proton}$ ~\cite{Touchard1982},
suggesting a large fractional component of single-particle 2s$_{1/2}$. 
The 80\% $\beta^-$ branch to the 1/2$^+$ 2.599 MeV state (Fig.~\ref{fig:47Kdecay}) has log(ft)=4.81, 
for which a literature shell-model calculation of the Gamow-Teller strength finds log(ft)$_{GT}$ = 4.39~\cite{Choudhary2021}. This experimental $g_A|M_{GT}|$=0.305 is considerably smaller than the single-particle 2s$_{1/2}$ $GT$ value of $\sqrt{3}$.  We include in our simulations the 18.4$\pm$0.3\% branch to the 2.578 MeV first excited 3/2$^+$ state, and another 1.3\% known to decay to five other 3/2$^+$ states~\cite{Smith2020}: these can have no Fermi component so simply dilute our measured $A_{\rm recoil}$.
For pure Gamow-Teller decay, the $\beta^-$ asymmetry $A_{\beta {GT}}$=-1/($I_i$+1)=-2/3 for the $1/2^+\rightarrow 1/2^+$ transition and $I_i/(I_i+1)$=+1/3 for the $1/2^+ \rightarrow 3/2^+$ transitions, so the precision of their
average $A_{\beta{GT}}$=-0.467 $\pm$ 0.020 is needed to extract $M_F$ from $A_{\beta}$.

Ref.~\cite{Smith2020} also observes a total of 0.042\% to first-forbidden branches. 
Although these can have asymmetries near unity, we can safely ignore them at our achieved accuracy.
The experimental upper limit for direct $\beta$ decay to the 2.013 MeV first 3/2$^-$ state would increase our uncertainty from branching ratios in Table~\ref{tab:table1} by a factor of 1.5. We constrain this branch with
a calculation of the first-forbidden nonunique fT value~\cite{Choudhary2021} of 0.25\%; our assignment of 0.25\% uncertainty does not perturb our uncertainty.


%
%

\section{Experiment: }


\subsection{\label{sec:level2}TRIUMF Neutral Atom Trap}

\begin{figure}[htb]
    \centering
    \includegraphics[width=0.70\linewidth]{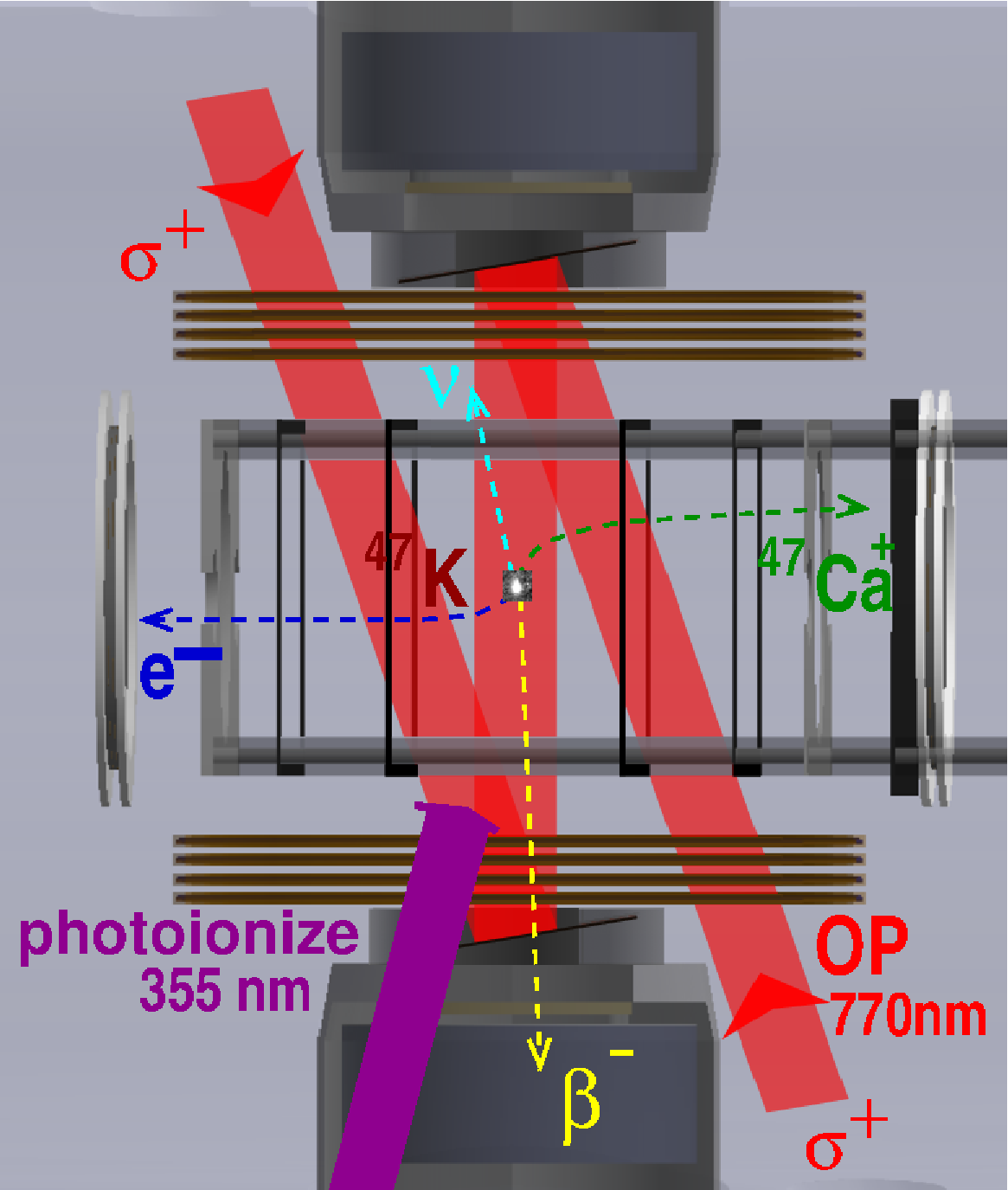}
    \caption{TRINAT during the optical pumping (OP) time. Shown are $\beta$ telescopes, mirrors for OP light and its beams, magnetic field coils, electric field electrodes, and microchannel plates for electron and ion detection. A CMOS camera image of 1,000 trapped atoms is superimposed. Distance between trap cloud and ion MCP is 9.7 cm.}
    \label{fig:trap}
\end{figure}

Fig.~\ref{fig:trap} is a side view of
the detection apparatus of  TRIUMF's Neutral Atom Trap for $\beta$ decay (TRINAT). Not shown is the collection trap from a vapor cell cube, nor the push beams between traps~\cite{Swanson1998}.

Using 6x10$^{6}$/s mass-separated $^{47}$K delivered from the TRIUMF-ISAC isotope separator on-line facility, we trapped on average 500-1,000 $^{47}$K atoms during the data-taking time. 
We optimized the number of atoms when the trapping light was tuned about 3 linewidths to the red of the 4S$_{1/2}$ to 4P$_{3/2}$ F=1 to F=2 transition, as measured with respect to the optical resonance measured by Ref.~\cite{Touchard1982}. Repumping light on the F=0 to F=1 transition was provided by 3.3 GHz fiber-coupled electro-optic modulators.


\subsection{Polarization by optical pumping}

The optical pumping scheme is similar to our $^{37}$K measurement~\cite{FenkerNJP2016}.
Changes include much thinner pellicle mirrors (4 $\mu$m polyimide with 100 nm Au coatings) along the optical pumping axis to reduce $\beta$ straggling.

  We alternate 2.9 ms trapping with 1.1 ms optical pumping, during which we make the polarized $\beta$ decay measurements. 
During the polarization time, we switch the magnetic field from the trap's quadrupole to a uniform 1 Gauss field pointed up, and apply circularly polarized light along the quantization axis.
      Once we start the OP cycle, atoms increase spin to maximum, then stop
      absorbing the S$_{1/2}$ to P$_{1/2}$ light. 
      If instead the light is linearly polarized, atoms keep absorbing, and the atoms and nuclei remain unpolarized.

      \begin{figure}[htb]
          \centering
          \includegraphics[width=0.65\linewidth]{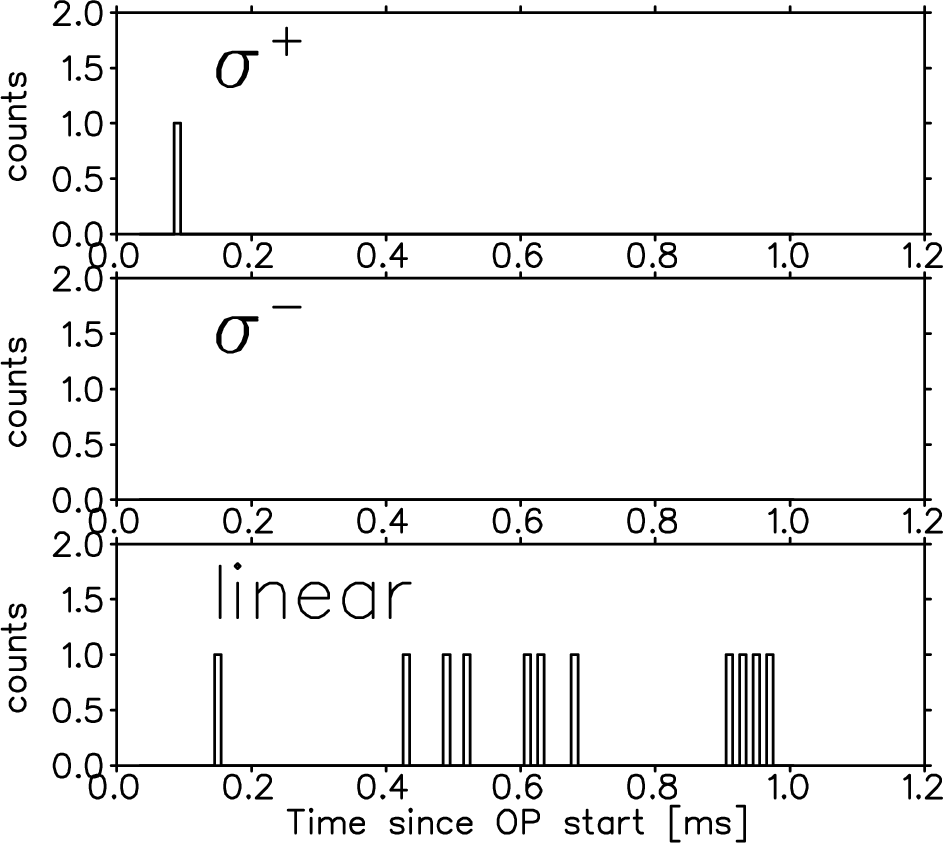}
          \caption{Excited state population during the optical pumping time for circularly and linearly polarized light. See text for deduction of nuclear polarization. }
          \label{fig:OP}
      \end{figure}

      Once excited,
      pulsed 355 nm light has enough energy/photon to photoionize from the $P$ states about 1/10$^6$ atoms per pulse, detected in the ion MCP.
      The photoions, distinguished by their time-of-flight (TOF) and centre position, determine average trap cloud sizes and positions when the magneto-optical trap (MOT) light is on. 
      
      The vanishing of photoion rate during the polarization time is then our probe of the  atomic polarization quality. 
                     We measured 11 photoions while linearly polarized (about 1/4 the total beamtime)  and
      1 photon circularly polarized (see Fig.~\ref{fig:OP}).
      The polarization extracted by rate equations from this circular/linear ratio is independent of power and detuning of the optical pumping light to well within this accuracy. 
      The fraction of nuclear polarization achieved for the decaying $^{47}$K atoms was $P$ = $\langle Iz \rangle/I$ = 0.96 $\pm$ 0.04.
%

\subsection{Geometry and Detectors}

An electric field collects $^{47}$Ca ions produced in $^{47}$K $\beta^-$ decay to an MCP with 78 mm active diameter located 9.7 cm away.
By comparing trajectories with a detailed finite element calculation, we found our simulations could assume a uniform 650 V/cm field.
Decay by $\beta^-$ naturally makes $^{47}$Ca$^{+1}$ ions.
  Additional low-energy atomic shakeoff electrons, which take an average of 6 ns to reach the opposite 40 mm diameter MCP, provide a starting trigger for the TOF of $^{47}$Ca$^{+2}$ and higher. 
%

         Critical to $\beta$ detection is discriminating  $\beta$'s
from $\gamma$'s, because their ratio in $^{47}$K decay is about 1 to 2. We use the same 0.30 mm thick
double-sided silicon strip detectors as Ref.~\cite{Fenker2018}, 
similarly requiring both X and Y strips above energy threshold and similar calibrated energy deposited.
Our plastic 4x9cm scintillators for $\beta^+$ detection now use silicon photomultiplier (SiPM) readout, characterized in Ref.~\cite{Ozen2023}. 

%
%
%

\section{Results}

\subsection{e$^-$--Recoil $^{47}$Ca Ion Coincidences}
\begin{figure}[htb]
    \centering
 \includegraphics[angle=90,width=0.9\linewidth]{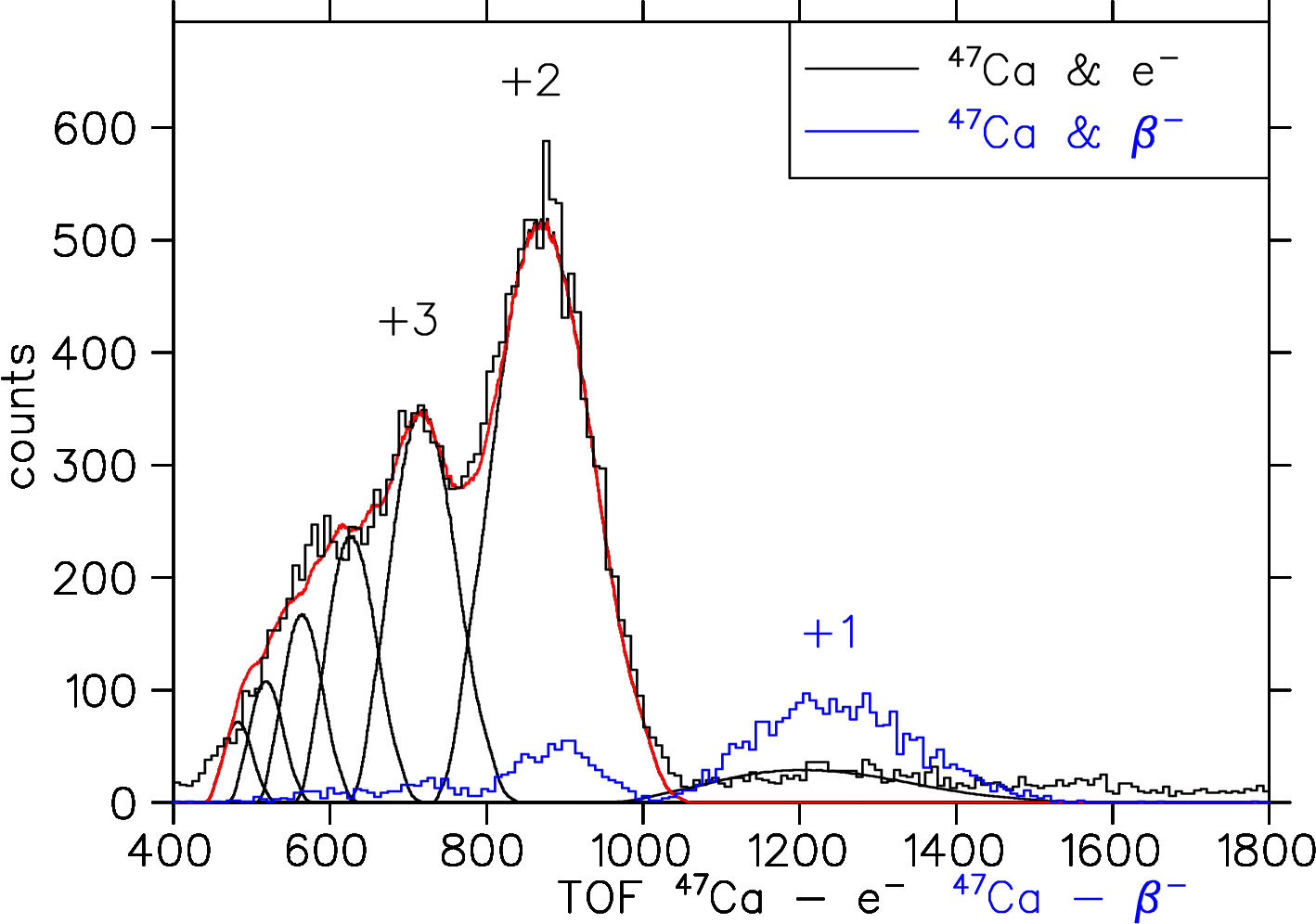} 
    \caption{Time-of-Flight (TOF) of $^{47}$Ca +2 to +7 ions started by shakeoff e$^-$, showing the modelled data decomposition. Blue histogram: TOF started by $\beta$'s in 
    the $\Delta E-E$ telescopes, which have lower statistics but less background from untrapped atoms and accidentals.}
    \label{fig:tof}
\end{figure}

\begin{figure}[htb]
    \centering
  \includegraphics[angle=0,width=0.9\linewidth]{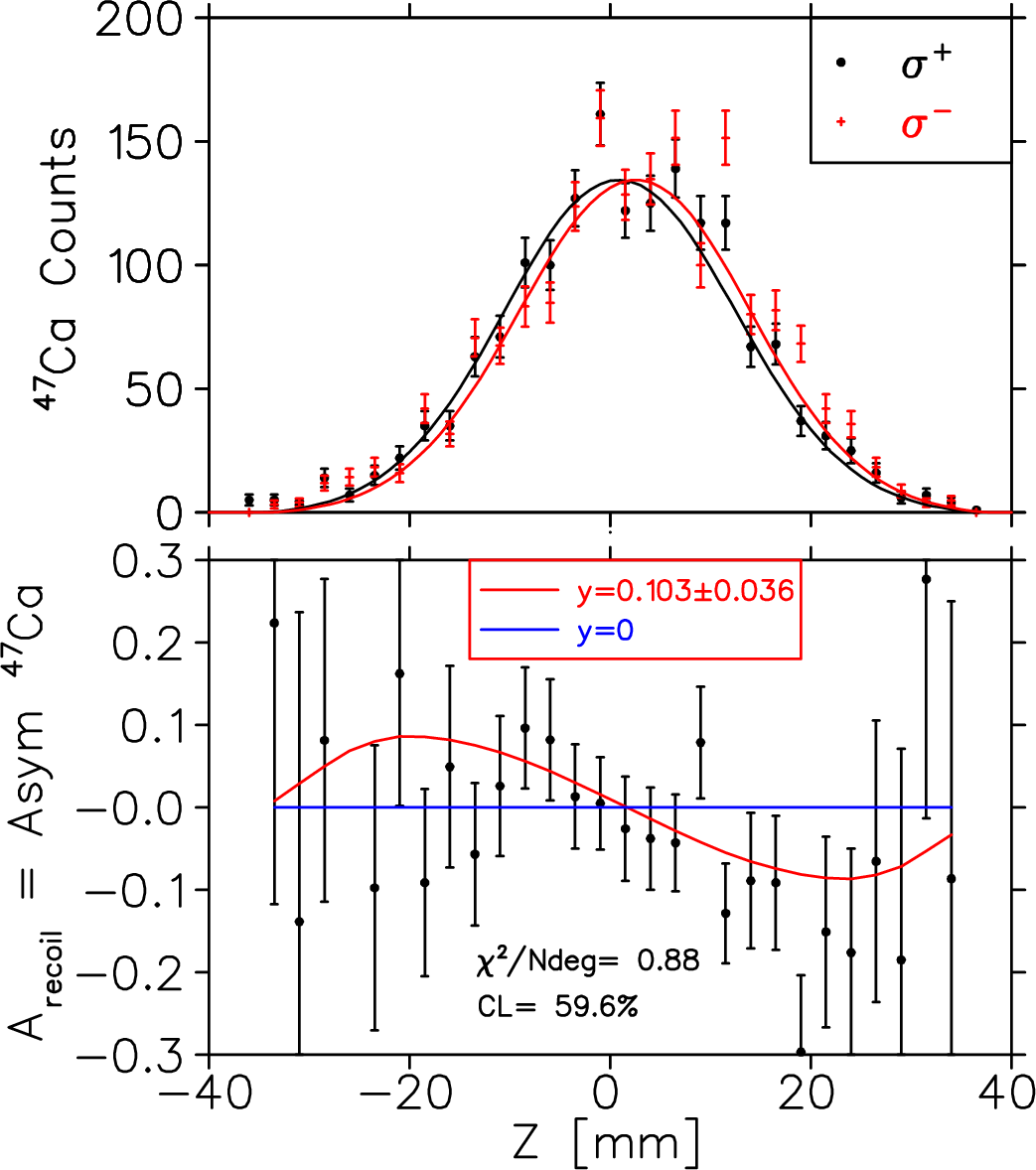}
    \caption{Top: Distribution along polarization axis Z of $^{47}$Ca$^{+2...7}$ in coincidence with shakeoff e$^-$'s, for the two polarizations. Bottom: The asymmetry of these distributions $A_{\rm recoil}$, i.e. the difference divided by the sum of the Top distributions. The nonzero asymmetry scales with $y$, and directly implies a nonzero Fermi contribution.}
    \label{fig:Arecoil}
\end{figure}

Our most sensitive channel detects 
coincidences between decay recoils on the ion MCP and shakeoff electrons on the e$^-$ MCP.
The TOF spectrum in Fig.~\ref{fig:tof} has contributions from $^{47}$Ca charge states +2 through +7. Their position asymmetry along the polarization axis is shown to be nonzero in Fig.~\ref{fig:Arecoil}, directly implying a nonzero Fermi contribution to the 1/2$^+$ $\rightarrow$ 1/2$^+$ transition.
 (Since $\beta^-$ decay makes +1 ions without shakeoff, the $e^-$ -- $^{47}$Ca events consistent with $^{47}$Ca$^{+1}$ TOF 
 are from $\beta$'s and $\gamma$'s firing the $e^-$ MCP and accidentals, so we do not use them for $A_{\rm recoil}$.)

We model this by a numerical integration of $\beta$ and $\nu$ (Eqs. 1-2), with the resulting $^{47}$Ca ions collected to the ion MCP by the uniform 650 V/cm electric field.
We find it adequate to include the momentum perturbation on the $^{47}$Ca from a single 2 MeV $\gamma$ subsequently emitted isotropically --- note emission of the dominant $\gamma$ in Fig.~1 must be isotropic. 
The result is $y$ = +0.103 $\pm$ 0.041 (a solution with $y \sim 10$ is physically excluded).  Systematic uncertainties are listed in Table~\ref{tab:table1}.

\begin{table}[hb]
\caption{\label{tab:table1}%
Systematic uncertainties. Both observables are statistics dominated. NA denotes not applicable. Listed common systematics are so different between observables that we are not concerned with their correlation.
}
\begin{ruledtabular}
\begin{tabular}{lll}
Source & $A_{\rm recoil}$ & pseudo$A_{\beta}$ \\
\colrule
        $A_{\rm recoil}$ bkg 6$\pm$4\%  &     0.014 & $<$ 0.002\\
	Polarization 0.96$\pm$0.04      &  0.004 & 0.023\\
 $\beta^-$ Branching ratio    &   0.002 & 0.022\\
   Weak magnetism   & 0.0006 & 0.0003\\
         Fit range in Z $\pm$20 to 34 mm   &        0.012 & NA\\
 $^{47}$Ca$^{+1}$ percent bkg & 0.001  & NA\\
 $^{47}$Ca$^{+N}$ distribution from TOF    &  $<$ 0.0005 & NA\\
	 E field     & negligible & 0.025 \\
   Backscatter correction -0.012$\pm$20\% & NA & 0.0024 \\
   \colrule
 Fit statistics & 0.037 & 0.082\\  
\colrule
        Total      &        0.041 &0.091
\end{tabular}
\end{ruledtabular}
\end{table} 


\subsubsection{Backgrounds from untrapped atoms} 
 The vacuum-limited trap half-life of 10 sec and the t$_{1/2}$=17.5 sec of $^{47}$K implies that more than half the nuclei decay after the atoms leave the trap. We have measured this background with 1 hour of data while deliberately ejecting atoms from the trap. We deduce a background of 6$\pm$4\% of the events in the e$^-$--$^{47}$Ca
 channel, roughly flat in TOF in the region we use of +2 through +7 charge states, and include that background in our simulation. This is consistent with a small fraction of the untrapped atoms sticking to the glassy carbon electrodes 
 close to the trap, while shakeoff 
 electrons from other surfaces are excluded from the electron MCP by the electric field.   Our $\beta$ collimation is sufficient that we see backgrounds consistent with zero for the $\beta$-recoil channel considered below.


\subsection{$\beta$--Recoil Coincidences Using Pseudo$A_\beta$}
We also measure $\beta$'s in coincidence with $^{47}$Ca recoils. If we measured $^{47}$Ca recoils over all directions and momenta, this would be a measurement of $A_{\beta}$. However, some
$^{47}$Ca$^{+1}$ ions with high transverse momenta do not impact the MCP, perturbing the asymmetry of $\beta$'s in coincidence by a well-defined combination of the $\beta$-$\nu$ correlation and the $\nu$ asymmetry.

\begin{figure}[htb]
    \centering
\includegraphics[angle=0,width=1\linewidth]{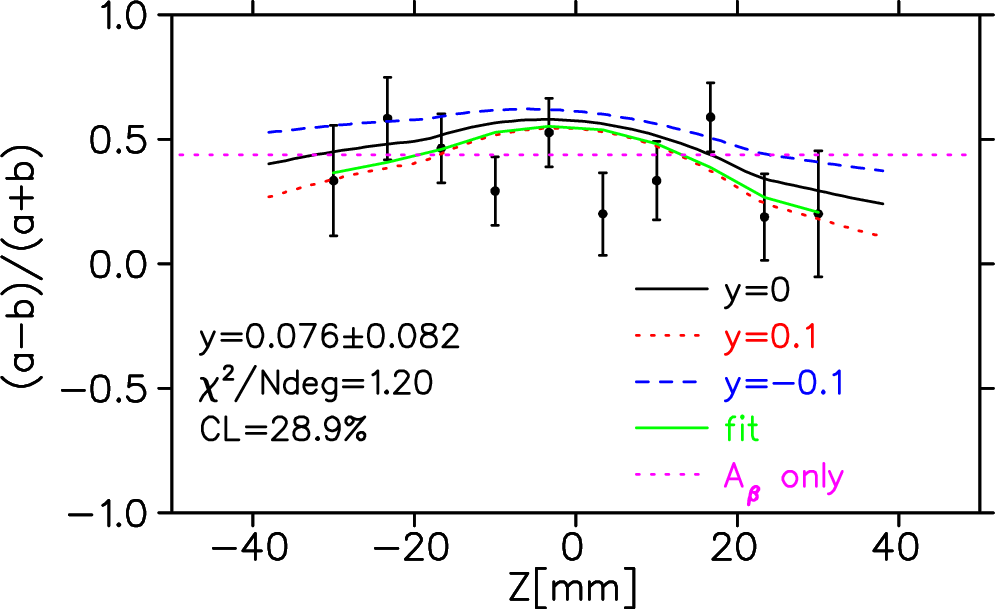}
     \caption{Similar to Fig.~\ref{fig:Arecoil}, but for $\beta$-$^{47}$Ca$^{+1}$ coincidences.}
   \label{fig:pseudoAbeta}
\end{figure}

This observable, which we name pseudo$A_{\beta}$, we also model by numerical integration, including the effects of a single 2 MeV $\gamma$.  We group the four combinations of $\beta$ detector and polarization sign in pairs to cancel asymmetries, denoted ``a'' and ``b'' in Fig.~\ref{fig:pseudoAbeta}.
 Numerical simulations for three values of M$_{F}$/M$_{GT}$ show sensitivity to the asymmetry, along with the best fit. A single straight line for M$_{F}$=0 and hypothetical full collection of $^{47}$Ca is shown to indicate how the asymmetries are distorted from $A_{\beta}$. 
 The significantly smaller  difference in asymmetry for positive vs. negative Z is due to 0.22 and 0.45 mm displacements in the cloud Z and horizontal position and subsequent change in $^{47}$Ca collection, and is well-reproduced by the simulation.

 We note that the sign of $A_{\beta}$ is determined from this observable. We use this to determine the sign of our spin polarization. 

To deduce
$y$ 
from pseudo$A_{\beta}$ requires more precision and accuracy than $A_{\rm recoil}$,
because of the nonzero $A_{\beta GT}$=-0.467.
The uncertainties are summarized in Table~\ref{tab:table1}. Based on our previous $^{37}$K $A_{\beta}$ measurement~\cite{Fenker2018}, we scale our experimental value by 1/1.023 to approximately account for backscatter, assigning here a more generous uncertainty of 20\% because we have not done full simulations of geometry changes.  

The result from pseudo$A_{\beta}$, 
$y$=0.076$\pm$0.091,
is consistent in sign with $A_{\rm recoil}$, but with larger uncertainty.


\subsubsection{Recoil order corrections.} We use Ref.~\cite{Holstein1974,Holstein1974erratum} for the correction from the recoil kinetic energy, the 1st-order in recoil correction from weak magnetism $b_W$, and the Coulomb finite-size correction. Assuming the wavefunction of initial and final 1/2$^+$ states is an s$_{1/2}$ nucleon, $b_W$ has no orbital angular momentum contribution and becomes the nucleon value~\cite{Hayes2017}: the first-class induced tensor $d_I$ also vanishes for s$_{1/2}$. Similarly assuming single-particle s$_{1/2} \rightarrow d_{3/2}$ for the 20\% branches to 3/2$^+$ states, the orbital correction to $b$ is zero if $l$ changes; here we ignore potentially nonzero $d_I$ and 2nd-order in recoil corrections because of the 20\% fraction.
The recoil corrections increase our deduced $A_{\rm recoil}$ by 0.0025, 0.0012 from $b_W$ to which we assign 50\% uncertainty; the Coulomb finite-size correction decreases $A_{\rm recoil}$ by 0.0017. Similarly, $b_W$ changes pseudo$A_{\beta}$ by 
0.0006$\pm$0.0003.

%
%

\subsection{Result and Isospin breaking}

Our weighted average of results from $A_{\rm recoil}$ and pseudo$A_{\beta}$ is then 
$y=g_V M_F/g_A M_{GT}$= 0.098$\pm$0.037 
for the 1/2$^+$ to 1/2$^+$ transition.

Given the measured $g_A M_{GT}$= 0.305, 
we deduce 
$|M_F|$= 0.030$\pm$0.011
(assuming $g_V=1.00$, and that we do not know the sign of $M_{GT}$).
To compare to other nuclei thought to be dominated by $\mathcal{A}-\bar{\mathcal{A}}$ mixing, 
we use a first-order perturbation theory expression
from the literature~\cite{Bhattacherjee1967,Bloom1964}: $M_F= \frac{\langle A | H_{\rm C} | \bar{A}\rangle }{\Delta E} \sqrt{(T \mp T_z)(T \pm T_z + 1)}$ 
(upper vs.\ lower sign for $\beta^-$ vs.\ $\beta^+$), along with the measured 
$\bar{\mathcal{A}}$-$\mathcal{A}$ splitting $\Delta E$ = 10.1 MeV~\cite{Burrows2007}, to deduce a Coulomb matrix element
$|H_{\rm C}|$ = 101 $\pm$ 37 keV.

This Coulomb matrix element is over half of the 160 keV prediction of $\mathcal{A}-\bar{\mathcal{A}}$ mixing.  We attribute this to the simple structure of nearly-doubly magic $^{47}$Ca and its single 1/2$^+$ state. That this is not the full prediction suggests the state is somewhat more complex than $\bar{\mathcal{A}}$.

\begin{figure}[htb]
    \centering
\includegraphics[angle=90,width=1\linewidth]{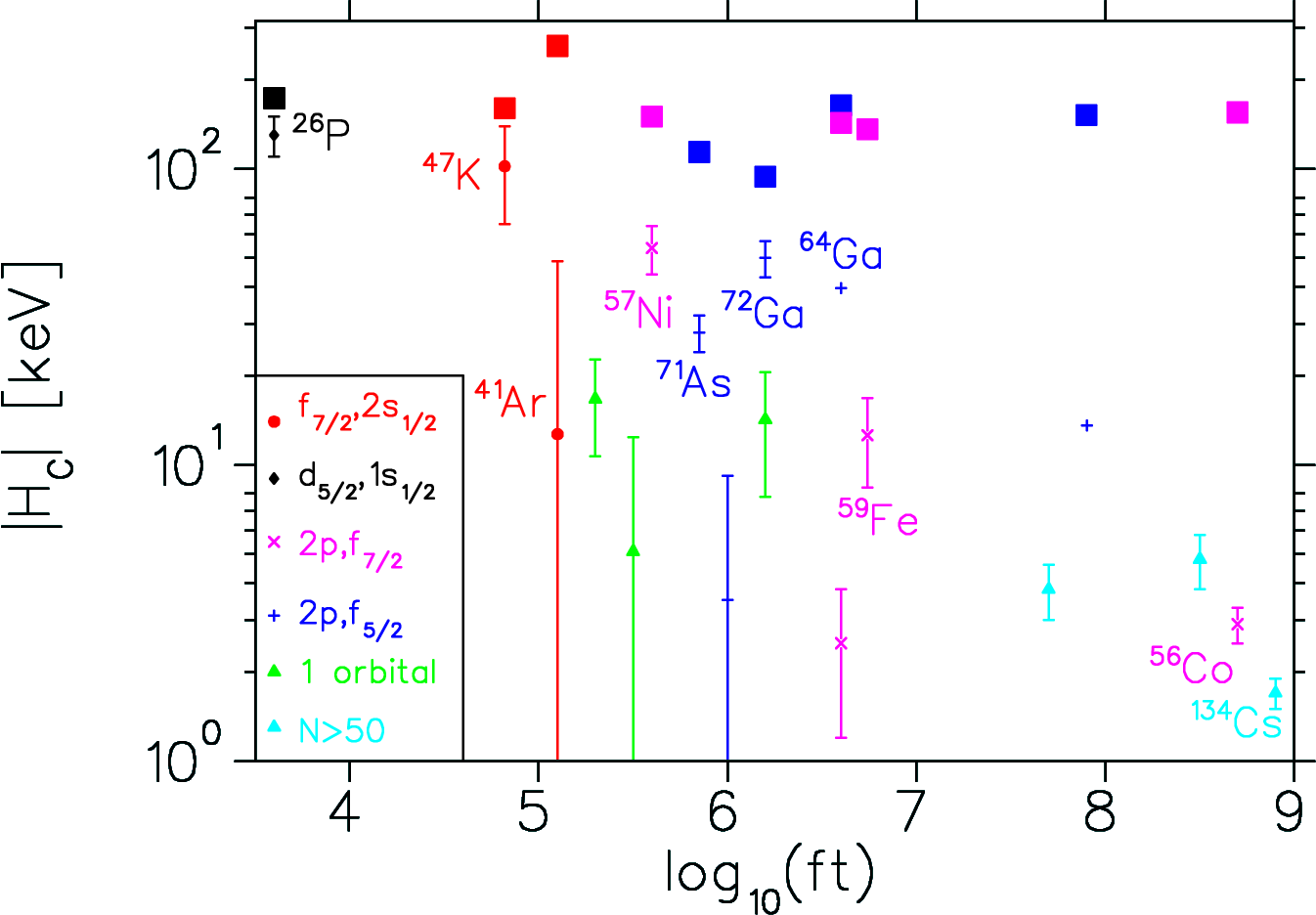}
\caption{
  Effective Coulomb mixing matrix element $H_C$ as a function of log(ft) for isospin-suppressed $\beta$ decay (Refs.~\cite{Mann1965,Atkinson1968,Behrens1967,Markey1982,Bhattacherjee1967,Severijns2005,Schuurmans2000,Liu2022}). 
Solid squares are $\mathcal{A}-\bar{\mathcal{A}}$ mixing from Eq.~2 for $^{47}$K decay's N=20 shell crossing, Eq.~1 for $^{26}$P assuming $d_{5/2} 1s_{1/2}$ excess proton occupancy, and the approximate use of Eq.~1 (i.e. Eq.~2 divided by 2) for all others~\cite{Auerbach2022} (see Section II.A). 
    } 
    \label{fig:Mf}
\end{figure}

Many $\beta$ decays in such systems have much smaller Coulomb matrix elements and $M_F$. Ref.~\cite{Auerbach2022} suggests the $\bar{\mathcal{A}}$ configuration is often fragmented among many states. 
Fig~\ref{fig:Mf} shows
measurements of $|H_C|$ with estimates from $\mathcal{A}-\bar{\mathcal{A}}$ mixing as a function of log(ft), a measure of nuclear complexity.
Although more complete calculations have difficulties reproducing its large $H_C$,
$^{26}$P~\cite{Liu2022} has a large fraction of 
Eq.~1's $\mathcal{A}-\bar{\mathcal{A}}$ prediction if $1d_{5/2}1s_{1/2}$ proton occupancy is naively assumed.
There are several cases with neutron excess in both 2p and 1f orbitals with relatively large $M_F$ and $H_{\rm C}$, factors of 2-4 smaller than their $H_{\rm C}$ predictions from Eq.~1~\cite{Auerbach2022}. 
In contrast, the $\beta^+$ decay of $^{56}$Co has a strikingly small $H_C$ given its relevant isobaric parent ($\mathcal{P}$) $^{56}$Fe also has excess neutrons spanning the N=28 shell closure: one explanation would be fragmentation of the $\bar{\mathcal{A}}$ configuration.
Cases with excess neutrons all in one orbital tend to have smaller $H_C$, consistent with little $\mathcal{A}-\bar{\mathcal{A}}$ mixing, though the most striking trend is the 
general drop of $|H_C|$ with $GT$ strength.

Ref.~\cite{Barroso1973} proposed a time-reversal measurement in $|y|$=0.21$\pm$0.01 $^{134}$Cs, which has a very small $M_{GT}$ and small Coulomb matrix element. A measurement was similarly pursued in $|y|$=0.13$\pm$0.02 $^{56}$Co~\cite{Calaprice1977}. Such time-reversal $\beta$ decay observables are proportional to $y$. 
For isovector TOPE nucleon-nucleon~\cite{Herczeg1966} matrix elments having similar dependence on nuclear complexity (e.g. a long-ranged isovector Yukawa could be similar to Coulomb), a faster decay like $^{47}$K with its large $\bar{\mathcal{A}}$ component and its sizable $y$ could
also be a favourable system for a time-reversal search.


%

\section{Conclusion}

For the $^{47}$K $\beta^-$  1/2$^+$$\rightarrow$1/2$^+$ transition, we have measured the ratio of Fermi to Gamow-Teller matrix elements 
$y$=0.098$\pm$0.037, and from the known $GT$ strength we deduce $|M_F|$=0.030$\pm$0.011.
Interpreted as 
$\mathcal{A}-\bar{\mathcal{A}}$
mixing in the progeny $^{47}$Ca, this result
implies a relatively large effective Coulomb mixing matrix element magnitude 
101$\pm$37 keV. A large matrix element of 160 keV is generated for $^{47}$Ca from 
$\mathcal{A}-\bar{\mathcal{A}}$
mixing, as its excess neutrons over protons occupy two major shells, f$_{7/2}$ and sd, with naturally different Coulomb energies~\cite{Auerbach2022}. The large fraction observed of that prediction we attribute to the existence of only one 1/2$^+$ state in the nearly doubly-magic $^{47}$Ca~\cite{Smith2020}.  

\begin{acknowledgments}

We acknowledge TRIUMF-ISAC staff, in particular for UC$_x$
target preparation. Supported by the
Natural Sciences and Engineering Research Council of
Canada and RBC Foundation,
TRIUMF
receives federal funding via a contribution agreement
through the National Research Council of Canada.
\end{acknowledgments}



\providecommand{\noopsort}[1]{}\providecommand{\singleletter}[1]{#1}%

\end{document}